\documentclass[english,aps,prl,twocolumn,superscriptaddress,showpacs]{revtex4}
\usepackage{graphicx}
\usepackage{babel}

\begin{document}

\title{Non-axisymmetric Anisotropy of Solar Wind Turbulence}

\author{A.J.~Turner}
\email{a.j.turner@warwick.ac.uk}
\affiliation{Centre for Fusion, Space and Astrophysics; University of Warwick,
Coventry, CV4 7AL, United Kingdom}

\author{G.~Gogoberidze}
\affiliation{Centre for Fusion, Space and Astrophysics; University of Warwick,
Coventry, CV4 7AL, United Kingdom}
\affiliation{Institute of Theoretical Physics, Ilia State University, 3/5 Cholokashvili ave., 0162 Tbilisi, Georgia}

\author{S.C.~Chapman}
\affiliation{Centre for Fusion, Space and Astrophysics; University of Warwick,
Coventry, CV4 7AL, United Kingdom}

\author{B.~Hnat}
\affiliation{Centre for Fusion, Space and Astrophysics; University of Warwick,
Coventry, CV4 7AL, United Kingdom}

\author{W.-C.~M\"uller}
\affiliation{Max-Planck-Institut f\"ur Plasmaphysik, Boltzmannstr. 2, 85748 Garching bei M\"unchen, Germany}

\begin{abstract}

 A key prediction of turbulence theories is frame-invariance, and in magnetohydrodynamic (MHD) turbulence, axisymmetry of fluctuations with respect to the background magnetic field.
 Paradoxically the power in fluctuations in the turbulent solar wind are observed to be ordered with respect to the bulk macroscopic flow  as well as the background magnetic field. Here, non-axisymmetry across the inertial and dissipation ranges is quantified using
  in-situ observations from Cluster. The
  observed inertial range non-axisymmetry is reproduced by a 'fly through' sampling of a Direct Numerical Simulation of MHD turbulence.
  Furthermore, 'fly through' sampling of a linear superposition of transverse waves with axisymmetric fluctuations generates the trend in non-axisymmetry with power spectral exponent.  The observed  non-axisymmetric anisotropy may thus simply arise as a
sampling effect related to Taylor's hypothesis and is not related to the plasma dynamics itself. 
\end{abstract}


\maketitle

Solar wind fluctuations observed by satellites in-situ
exhibit power law scaling regions identified with an inertial range of magnetohydrodynamic (MHD) turbulence,
 and with a 'dissipation' range below ion kinetic scales, providing a natural laboratory for plasma turbulence
  (for a recent review see, e.g., Ref. \cite{BC05}).
In hydrodynamic turbulence, any anisotropy in fluctuations at large scales will
tend to isotropize as the cascade proceeds to smaller scales \cite{MY}.
The situation is different in plasma turbulence where the existence of a mean magnetic field
sets a natural preferential direction for anisotropy.
Anisotropy is thus a key topic in
theoretical \cite{GS95,B06,G07}, numerical \cite{MG01,MBG03}, and observational studies of plasma turbulence
in the solar wind \cite{BD71,CH07,NCD09,NEa10,SEa10,WEa11}.

The seminal study of Belcher and Davis \citep{BD71} used Mariner
5 observations to investigate anisotropy of the solar wind magnetic fluctuations in the low frequency
(energy containing) and inertial intervals.
They found that the fluctuations on average have $5:4:1$ power anisotropy in
an orthogonal coordinate system whose axis are [$\mathbf{e}_{B}\times \mathbf{e}_{R},\mathbf{e}_{B} \times (\mathbf{e}_{B} \times \mathbf{e}_{R}),\mathbf{e}_{B}$], where
$\mathbf{e}_{B}$ is a unit vector in the average magnetic field direction and $\mathbf{e}_{R}$
is a unit vector radially away from the sun. This conclusion, that solar wind
fluctuations are non-axisymmetrically anisotropic with respect to the magnetic field direction in the low frequency and
inertial intervals was confirmed by different authors \cite{BS75}. Recent results using $k$-filtering  \cite{NEa10} were consistent with the results of  \cite{BD71} in that the main power was found in the plane perpendicular to the local magnetic field distributed preferentially  in the direction perpendicular to both the magnetic field and the solar wind velocity.
Dissipation range magnetic fluctuations have also been found to be non-axisymmetric
 \cite{PEa09} using minimum variance analysis \cite{SC67}.

The anisotropic expansion of the solar wind can introduce a preferred direction as captured by models \cite{Volk,Grappin} and as observed on longer timescales (5-12 hrs, see \cite{Saur}). However, from the perspective of turbulence, ordering of the observed non-axisymmetric power anisotropy with the direction of the  solar wind bulk flow velocity at the inertial and dissipation scales is rather unexpected. If the macroscopic bulk flow speed is sufficiently large compared to
 that of the fluctuations and that of the characteristic wave speeds of the plasma, then on the timescales over which we observe
 turbulence in-situ this bulk flow simply acts
 to advect the fluctuating plasma. This is Taylor's hypothesis \cite{Taylor}, and if it holds, then since the observed properties of the evolving turbulence
 are frame independent they should not correlate with the macroscopic flow direction. The observation
of non-axisymmetry \cite{BD71,NEa10}  with respect to the macroscopic flow direction in the high speed solar wind flow is thus
paradoxical.
 In plasma turbulence,  one would anticipate
axisymmetric anisotropy ordered with respect to the local magnetic field.
Indeed, theories of MHD and kinetic range turbulence assume
axisymmetry of statistical characteristics \cite{GS95,B06,G07,VG05,TEa09} (note that although the model
developed in Ref. \cite{B06} implies {\it local} non-axisymmetry of turbulent eddies, it
still assumes an axisymmetric energy spectrum). As a consequence,
studies of anisotropy of  solar wind turbulence
using single spacecraft observations often assume axisymmetry
(see, e.g., \cite{CH07,CEa10,WEa11} and references therein).
Understanding the origin of this non-axisymmetry is the subject of this Letter, and is essential if
solar wind observations are to be employed in the study of turbulence, in particular in the context of direct
comparisons  between theoretically predicted and
observed statistical properties and scaling exponents.

Here, we show that the observed non-axisymmetric anisotropy can arise as a data sampling effect rather than as a physical property of the turbulence.
We first sample the output of a direct numerical simulation (DNS) of MHD turbulence with a 'fly through' emulating single spacecraft in-situ observations using Taylor's hypothesis. We will see that this is sufficient to reproduce the observed non-axisymmetry in the inertial range of the solar wind. To understand how this non-axisymmetry can arise, we  consider the simplest scenario- a 'fly through' sampling of a fluctuating field composed of linearly superposed transverse waves with axisymmetric power anisotropy. The only free parameter in this model is the power spectral exponent of perpendicular fluctuations. This model reproduces the observed trend - that the non-axisymmetry increases with the perpendicular power spectral exponent as we move from the inertial to the dissipation range of scales.
 The observed  non-axisymmetric anisotropy may thus simply arise as a
sampling effect related to Taylor's hypothesis.

We present the analysis of a sample interval [January 20, 2007, 1200-1315 UT] of fast quiet solar wind
observed by Cluster spacecraft 4 whilst the magnetic field instruments FGM and STAFF-SC were in burst mode, providing a simultaneous observation across the inertial and dissipation ranges.
FGM (sampled at $67~{\rm Hz}$) and  STAFF-SC data (sampled at $450~{\rm Hz}$) are combined  by the same procedure as in  \cite{CEa10,A04}, where a discrete wavelet transform is applied to both instrument data sets. This merging procedure generates one time series
containing frequencies ranging from the highest frequency of the STAFF-SC data and the lowest frequency of the FGM data.
This interval is of fast solar wind with a flow speed of
$\sim 590~{\rm km/s}$ with plasma parameters: average magnetic field $\overline{B}\simeq4~{\rm nT}$,
proton plasma $\beta\simeq{\rm 1.5}$, proton density $\rho_p \simeq~2~{\rm cm^{-3}}$,
proton temperature $T_{p}\simeq29~{\rm eV}$ and Alfv\'{e}n speed $V_{A}\simeq60~{\rm km/s}$.

We use the continuous wavelet transform (CWT), as outlined in  \cite{P09,H08}, to select fluctuations
on a specific scale, $\tau$. The fluctuations are resolved at each $\tau$ by a CWT performed on each
component of the magnetic field data, $\textbf{B}(t_{j})$, using the Morlet wavelet, to give
a fluctuation vector, $\delta\textbf{B}(t_{j},\tau)$. The vector fluctuation  are then projected onto the local field. At each scale $\tau$
the local magnetic field is defined for every time, $t_{j}$, by the convolution of a Gaussian window
of width $2\tau$ centred on $t_{j}$ with the data, such that
$
\mathbf{\overline{B}}(t_{j},\tau) = [\mathbf{B}(t_{j}) \ast g(t_{j},\tau)]$
where $g$ is the Gaussian window. The scale
$\tau$ is related to frequency $f$, (in Hz) of the central frequency of the Morlet wavelet.
 This allows the local magnetic field and the fluctuations to be rewritten as a functions
 of time and frequency $\overline{\textbf{B}}(t_{j},f)$ and $\delta\textbf{B}(t_{j},f)$, respectively.

We define the local system of unit vectors following \cite{BD71}. The unit vector in the direction of the local magnetic field is
$\mathbf{e}_{z}(t_{j},f)=\overline{\mathbf{B}}(t_{j},f)/\left|\overline{\mathbf{B}}(t_{j},f)\right|$.
The other two perpendicular unit vectors are ordered with respect to the macroscopic flow velocity direction, 
such that
\begin{equation}
\mathbf{e}_{x}(t_{j},f)= \frac{ \mathbf{e}_{z}\times \hat{\mathbf{V}}}{ \left|\mathbf{e}_{z}\times \hat{\mathbf{V}} \right|},~~
\mathbf{e}_{y}(t_{j},f)=\mathbf{e}_{z}\times\mathbf{e}_{x}, \label{eq:e_x}
\end{equation}
where $\hat{\mathbf{V}}$
is the unit vector of the bulk flow velocity
direction  time-averaged over the interval.
 During this interval the spacecraft is in a fast and steady stream of the solar wind,
thus the local velocity is close to the time averaged macroscopic  velocity in (\ref{eq:e_x}). 

It can be shown that the results obtained are not sensitive to the small level of observed variations in the solar wind direction. The magnetic field fluctuations are then projected to the new basis (\ref{eq:e_x}) and the Power Spectral Density (PSD) of the corresponding components are defined by:
\begin{equation}
PSD_{z,x,y}(f)=\frac{2\Delta}{N}\sum_{j=1}^{N}\delta B_{z,x,y}(t_{j},f)^{2}\,\label{eq:1}
\end{equation}
where $N$ is the sample size at each frequency and $\Delta$ is sampling interval of the data.

\begin{figure}
\begin{centering}
\includegraphics[width=1\columnwidth]{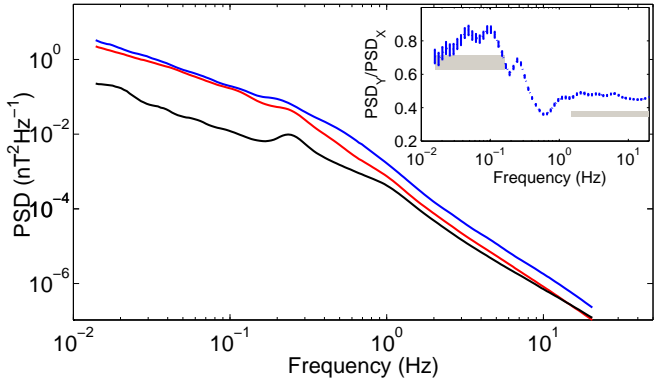}
\par\end{centering}
\caption{$PSD$s of magnetic field components (from bottom to top) - $PSD_z$ (black), $PSD_y$ (red) and
$PSD_{x}$ (blue), where $\mathbf{e}_{z}= \overline{\mathbf{B}}/\left|\overline{\mathbf{B}}\right|, ~\mathbf{e}_{x}= \mathbf{e}_{z}\times \hat{\mathbf{V}}/ \left|\mathbf{e}_{z}\times \hat{\mathbf{V}} \right|$ and $\mathbf{e}_{y}=\mathbf{e}_{z}\times\mathbf{e}_{x}$. Error bars are always smaller than $4\% $ and are usually
smaller than the line width. In the insert - The ratio of perpendicular PSDs $PSD_{y}/PSD_{x}$.} \label{fig:PSD}
\end{figure}

We plot the PSD of these components for this Cluster interval in Fig. \ref{fig:PSD}.
In the Figure, $PSD_z$, $PSD_y$ and $PSD_{x}$  are (from bottom to top) represented by black, red and
blue lines, respectively.
The frequency range captured by this interval of data  covers both the dissipation range and the high frequency
part of the inertial range.
 The average ratio of the total perpendicular
to the parallel PSD in the inertial range  does not vary significantly with frequency and is $\sim 17:1$. This is consistent with prior observations \cite{BS75} that the power in the inertial range is predominately perpendicular to the local magnetic field, implying Alfv\'enic fluctuations. In the dissipation range the power isotropises \cite{Leamon,Kiyani}. The ratio of the PSD of the perpendicular components is shown in the inset of the Figure.
 The mean value of the
ratio of the PSD of the perpendicular components in the inertial range is $PSD_{y}/PSD_{x}\simeq 0.8$, in close
agreement with \cite{BD71}. This ratio increases significantly in the dissipation range  to  $PSD_{y}/PSD_{x}\simeq 0.45$, coincident with a steepening of the PSD power spectral exponents. We will explore the origin of these average values and trends. Ratios of the PSD can emphasize  fluctuations in the PSD that are localized in frequency, these can be observational artefacts such as spacecraft spin tones, or can be of physical origin such as power enhancements at the cross-over from the inertial to dissipation range. The PSD ratio can be seen to oscillate at the transition between the inertial and dissipation range due to these effects.
We performed the above PSD analysis described on a second Cluster interval; a
fast stream (January 30, 2007, 0000-0100 UT) and obtained similar results.

\begin{figure}
\begin{centering}
\includegraphics[width=1\columnwidth]{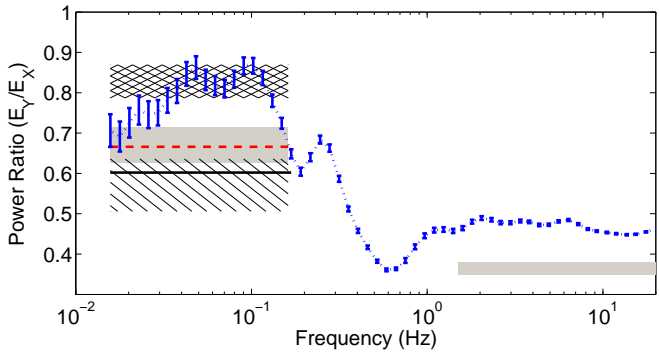}
\par\end{centering}
\caption{The ratio $E_{y}/E_{x}$, where $\mathbf{e}_{x}$ and $\mathbf{e}_{y}$ refer to a field aligned coordinate system as in Figure \ref{fig:PSD}, for the solar wind data (blue dotted line), the range of values obtained from a number of simulation 'fly throughs' for Case I (crosshatched area) and Case II (hatched area), as well as predictions of Eq. (\ref{eq:6}) for the spectral indices
$\gamma=5/3$ (black solid line), $\gamma=1.5$ (red dashed line), $\gamma =1.5 \pm 0.1$ (left gray area) and $\gamma=2.76 \pm 0.05$ (right gray area).} \label{fig:Per_Rat}
\end{figure}

We now see how this non-axisymmetry can arise.   Spatial snapshots of the three-dimensional velocity and magnetic fields of developed incompressible MHD turbulence provide a model for the inertial range fluctuations, these were obtained from the DNS described in \cite{MBG03}. We will consider two simulations: Case I is a globally isotropic freely decaying turbulence with resolution $512^3$ and Case II corresponds to $1024^2 \times 256$ forced turbulence simulation with strong background magnetic field $B_0$, such that $b_{rms}/B_0 \sim 0.2$ \cite{MBG03}.  A 'fly through' sampling is performed on these spatial snapshots by sampling a 'time series' of one-dimensional data along different straight line paths through the simulation domain at constant velocity to mimic single satellite observation of the turbulence using Taylor's hypothesis. The direction of the straight line path defines the pseudo- macroscopic flow direction $\hat{\mathbf{V}}$ in the components defined in (\ref{eq:e_x}).
From each 'fly through'  we can then calculate the time averaged ratio of the power in the two  perpendicular components in the same manner as above for the solar wind data. These simulations offer two contrasting, and informative, numerical experiments.
In Case I the  background field direction is free to vary and a wide range of angles between the local background field and the pseudo-macroscopic flow can be realized.
 For Case II the 'fly through' is restricted to the plane perpendicular to the applied background magnetic field, to ensure that the \emph{local} background magnetic field is nearly perpendicular to the pseudo-macroscopic flow direction- this will provide us with a simple test case in what follows. The results are presented in Figure \ref{fig:Per_Rat}. The blue dotted line corresponds to the solar wind data shown in the inset of Figure 1. The crosshatched and hatched areas indicate the range of values obtained from a number of simulation 'fly throughs' for Case I and Case II, respectively.
  Looking at the inertial range, we then see a remarkable agreement between the average of the power ratio in the inertial range in the solar wind, and Case I. Case II also shows non-axisymmetry but tends to overemphasize the value as compared to that observed by Cluster in the solar wind.

A full understanding of the origin of power non-axisymmetry in the solar wind should capture both  that it is seen in the inertial and dissipation range of scales and that it is strongest in the dissipation range. We now show that these features can be qualitatively captured by quite simple considerations.
 Consider a linear superposition of waves transverse to a constant background magnetic field ${\bf B}_0$.  For simplicity we assume ${\bf V}_{sw} \perp {\bf B}_0$. To fix coordinates,
 the solar wind velocity ${\bf V}_{sw}$ and ${\bf B}_0$ are directed along $y$ and $z$, respectively so that  the magnetic
fluctuations $\delta {\bf B}$ associated with the waves are in the $x,y$ plane. If the Fourier
amplitudes of the fluctuations are $\delta {\bf B}({\bf k})$ then under Taylor's hypothesis
the energy densities $E_y(\omega)$ and $ E_{x}(\omega)$ observed for the components parallel and perpendicular to ${\bf V}_{sw}$ are given
by (see also \cite{Forman})
\begin{eqnarray}
E_{y,x}(\omega)= \frac{1}{8 \pi} \int {\rm d}^3{\bf k} | \delta {\bf B}_{y,x}({\bf k}) |^2 \delta(\omega-{\bf k} \cdot {\bf V}_{sw}).
\end{eqnarray}
If we now assume  that there exists some scaling relation between the parallel and perpendicular wave numbers
(for example,  critical balance \cite{GS95}) and given $\delta {\bf B}({\bf k}) \perp {\bf k}$, after integration over ${\bf k_\parallel}$
we obtain
\begin{eqnarray}
E_{y}(\omega)=\int {\rm d}^2{\bf k}_\perp E_{2D}({\bf k}_\perp)  \delta(\omega-k_y V_{sw}) \sin^2\alpha, \label{eq:4}\\
E_{x}(\omega)=\int {\rm d}^2{\bf k}_\perp E_{2D}( {\bf k}_\perp)  \delta(\omega -k_y V_{sw}) \cos^2\alpha,\label{eq:5}
\end{eqnarray}
where $E_{2D}( {\bf k}_\perp)$ is two-dimensional
spectrum of fluctuations \cite{GS95} and $\alpha$ is the angle between ${\bf k}_\perp$ and ${\bf V}_{sw}$. Expressions (\ref{eq:4}) and (\ref{eq:5}) generally integrate to give $E_{y}(\omega)\neq E_{x}(\omega)$. In particular, for axi-symmetric fluctuations $E_{2D}( {\bf k}_\perp)=E_{2D}( k_\perp) = C k_\perp^{-\gamma-1}$, where $\gamma$ corresponds to a one-dimensional spectrum, Eqs. (\ref{eq:4}-\ref{eq:5}) yield
\begin{eqnarray}
E_{y}(\omega)=C \frac{\omega^2}{V_{sw}^2} \int \left(\frac{\omega^2}{V_{sw}^2}+k_x^2\right)^{-\frac{\gamma-3}{2}}  {\rm d} k_x, \\
E_{x}(\omega)=C \int k_x^2  \left(\frac{\omega^2}{V_{sw}^2}+k_x^2\right)^{-\frac{\gamma-3}{2}} {\rm d} k_x. 
\end{eqnarray}
This can be
integrated in terms of Beta functions and after straightforward manipulation (for $\gamma>0$) we obtain
\begin{eqnarray}
E_{y}(\omega)/E_{x}(\omega)=1/\gamma. \label{eq:6}
\end{eqnarray}
Thus the ratio $E_{y}(\omega)/E_{x}(\omega)$ is a decreasing function of the spectral index in qualitative agreement with the results presented above.
The values of the ratio $E_{y}(\omega)/E_{x}(\omega)$ for different values of $\gamma$ are indicated on Figure 2. Specifically, for $\gamma =1.5 \pm 0.1$ (inertial range spectral index observed here) $E_{y}(\omega)/E_{x}(\omega)=0.67 \pm 0.04$ (gray area in the low frequency range), for $\gamma=5/3$ (Kolmogorov spectrum) $E_{y}(\omega)/E_{x}(\omega)=0.6$ (black solid line), for $\gamma=3/2$ (Iroshnilov-Kraichnan spectrum) $E_{y}(\omega)/E_{x}(\omega)=0.67$ (red dotted line) and for $\gamma=2.76 \pm 0.05$ (dissipation range spectral index observed here) $E_{y}(\omega)/E_{x}(\omega)=0.36 \pm 0.02$ (gray area in the high frequency range).

This result has a simple physical explanation. When the
satellite samples three-dimensional fluctuations as a one-dimensional series all the fluctuations with
fixed $k_y$ contribute to a given frequency $\omega=k_y V_{sw}$. This mixture of different modes does not produce
any non-axisymmetric anisotropy if energy density is equally distributed among different scales ($\gamma=1$), but if this is not the case, the 
resultant frequency spectra for the energy density of different components will be non-axisymmetric. Indeed, consider the case
when energy density is decreasing with scale ($\gamma>1$). Then, at a given frequency $f=k_y V_{sw}$, since the waves are transverse,
modes with $k_x<k_y$ contribute more to  $E_{x}(\omega)$ compared to $E_{y}(\omega)$. For the modes with $k_x>k_y$
the situation is reversed, but because the modes in the latter range have less power, one finally has $E_{x}(\omega) > E_{y}(\omega)$.

Results from the simple model can be seen from Figure 2 to be consistent with the results from Case II of the DNS which share the same restricted geometry, that is, the background field is perpendicular to the macroscopic flow velocity.
The model also gives the qualitative trend in the anisotropy with  the spectral index of the turbulence. On the other hand, the simple model predicts stronger power anisotropy than seen in the solar wind as well as in the DNS without strong background magnetic field (Case I). This suggests that the overestimation in non-axisymmetry is indeed due to this restricted geometry of the simple model. To practically test this idea would require knowledge of the distribution of the angle between ${\bf V}_{sw}$ and ${\bf B}$ which, within any given interval of solar wind data is non-uniform, and which can vary from one interval to another. 

The sampling effects described above have implications for the results
obtained by methods used for the study of turbulence anisotropy that rely upon Taylor's hypothesis.  In addition, multispacecraft methods that do not
 use Taylor's hypothesis explicitly, such as 
 $k$-filtering (see, e.g., \cite{NEa10} and references therein) still require
filtering of the data in the spacecraft frame - selecting  perturbations within some frequency interval
$[-\omega_{max},\omega_{max}]$. As the solar wind speed is much larger than the  Alfv\'en speed, it is clear that
this procedure is essentially a filtering with respect
to the wave vector component parallel to the solar wind velocity. As demonstrated here, this leads to preferential filtering
of the modes with wave vectors mainly parallel to the solar wind, and as a result could generate non-axisymmetric anisotropy.

Finally, we have for the restricted case where the background field is perpendicular to the flow, obtained an explicit relationship between the spectral exponent and the power non-axisymmetry. For this special case, the non-axisymmetry in power is quite sensitive to the value of the exponent. It may be possible to develop this formalism, at least numerically, to use the observed non-axisymmetry in power to determine the scaling exponent more accurately. This is central to testing predictions of turbulence theories.

\begin{acknowledgments}
The authors acknowledge the Cluster instrument teams for providing FGM and STAFF-SC data. This work was supported by the UK STFC.
\end{acknowledgments}

\newpage

\begin{thebibliography}{}
\bibitem{BC05} R. Bruno and V. Carbone, Living Rev. Sol. Phys. {\bf 2}, 4 (2005).
\bibitem{MY} A.S.~Monin and A.M.~Yaglom, {\sl Statistical Fluid Mechanics}
(MIT Press, Cambridge, MA, 1975).
\bibitem{GS95} P.~Goldreich and S.~Sridhar, Astrophys. J. {\bf 438}, 763 (1995).
\bibitem{B06} S.~Boldyrev, Phys. Rev. Lett. {\bf 96}, 115002 (2006).	
\bibitem{G07} G.~Gogoberidze, Phys. Plasmas {\bf 14}, 022304 (2007).		
\bibitem{MBG03} W.-C.~Mueller and R.~Grappin, Phys. Rev. lett. {\bf 95}, 114502 (2005).	
\bibitem{MG01} J.~Maron and P.~Goldreich, Astrophys. J. {\bf 554}, 1175 (2001).
\bibitem{BD71} J.W.~Belcher and L.~Davis, Astrophys. J. {\bf 76}, 3534 (1971).
\bibitem{CH07} S. C. Chapman, B. Hnat, Geophys. Res. Lett., {\bf 34}, L17103, (2007).
\bibitem{NEa10} Y.~Narita et al., Phys. Rev. Lett. {\bf 104}, 171101 (2010).
\bibitem{SEa10} F.~Sahraoui et al., Phys. Rev. Lett. {\bf 105}, 131101 (2010).
\bibitem{WEa11} R.T.~Wicks et al., Phys. Rev. Lett. {\bf 106}, 045001 (2011).
\bibitem{NCD09} R.M.~Nicol, S.C.~Chapman, and R.O.~Dendy, Astrophys. J. {\bf 703}, 2138 (2009).
\bibitem{BS75} S.C.~Chang and A.~Nishida, Astrophys. Space Sci. {\bf 23}, 301 (1973);
J.W.~Belcher and C.V.~Solodyna, J. Geophys. Res. {\bf 80}, 181 (1975);
L.F.~Burlaga and J.M.~Turner, J. Geophys. Res. {\bf 81}, 73 (1976);
B.~Bavassano, et al., Solar Phys. {\bf 78}, 373 (1982);
E.~Marsch and C.-Y. Tu, J. Geophys. Res. {\bf 95}, 8211 (1990);
L.W.~Klein, D.A.~Roberts and M.L.~Goldstein, J. Geophys. Res. {\bf 96}, 3779 (1991).
\bibitem{PEa09} S.~Perry et al., J. Geophys. Res. {\bf 76}, A02102 (2009).
\bibitem{SC67} B.U.O.~Sonnerup, L.J.~Cahill, J. Geophys. Res. {\bf 72}, 171 (1967).
\bibitem{Volk} J.H~Volk and W.~Aplers, Astrophys. Space. Sci. {\bf{20}}, 267 (1973).
\bibitem{Grappin} R.~Grappin, M.~Velli and A.~Mangeney, Phys. Rev. Lett. {\bf{70}}, 2190 (1993)
\bibitem{Saur} J.~Saur and J.W.~Bieber, J. Geophys. Res. {\bf{104}}, 9975 (1999).
\bibitem{Taylor}G. I. Taylor, Proc. Lond. Math. Soc.,  {\bf 20} ,  196, (1921).
\bibitem{VG05} Y.~Voitenko and M.~Goossens, Phys. Rev. Lett. {\bf 94}, 135003 (2005).
\bibitem{TEa09} T.~Tatsuno et al., Phys. Rev. Lett. {\bf 103}, 015003 (2009).
\bibitem{CEa10} C.H.K.~Chen et al., Phys. Rev. Lett. {\bf 104}, 255002 (2010).
\bibitem{A04} O.~Alexandrova et al., J. Geophys. Res. {\bf 109}, A05207 (2004).
\bibitem{P09} J.J.~Podesta, Astrophys. J. {\bf 698}, 986 (2009).
\bibitem{H08} T.S.~Horbury, M.~Forman and S.~Oughton, Phys. Rev. Lett. {\bf 101}, 175005 (2008).
\bibitem{Leamon} R.J.~Leamon at al., J. Geophys. Res. {\bf{103}}, 4775 (1998).
\bibitem{Kiyani} K.H~Kiyani et al., Phys. Rev. Lett. {\bf 103}, 075006 (2009).
\bibitem{Forman} M.A.~Forman, R.T~Wicks and T.S.~Horbury, Astrophys. J. {\bf 733}, 76 (2011).


\end{thebibliography}
\end{document}